\documentclass[aps,pra,twocolumn,superscriptaddress,amsmath,amssymb,preprintnumbers,showpacs]{revtex4}
\pdfoutput=1
\usepackage{epsfig}
\usepackage{graphicx}
\usepackage{verbatim}
\usepackage{times}

\begin{document}

\newcommand{\re}[1]{\mathrm{Re}\left[#1\right]}
\newcommand{\im}[1]{\mathrm{Im}\left[#1\right]}
\newcommand{\abs}[1]{\left\vert#1\right\vert}
\newcommand{\nt}[1]{#1_{\mathbf{k}}^p}
\newcommand{\p}[2]{#1_{\mathbf{k}}^{#2}}
\newcommand{\po}[1]{#1_{\mathbf{k}}}
\newcommand{\arctanh}[0]{\text{arctanh}}
\newcommand{\arccoth}[0]{\text{arccoth}}
\newcommand{\arcsinh}[0]{\text{arcsinh}}
\newcommand{\arccosh}[0]{\text{arccosh}}
\newcommand{\unit}[0]{1}
\newcommand{\todo}[1]{{\small\tt[#1]}}

\newcommand{\prefactor}{\aleph} 
\newcommand{\halfprefactor}{\frac{ \aleph }{ 2 }} 

\title{The Role of Surface Plasmons in The Casimir Effect}

\author{F. Intravaia}
\affiliation{Institut f\"{u}r Physik, Universit\"{a}t Potsdam, Am 
Neuen Palais 10, 14469 Potsdam, Germany}
\author{C. Henkel}
\affiliation{Institut f\"{u}r Physik, Universit\"{a}t Potsdam, Am 
Neuen Palais 10, 14469 Potsdam, Germany}
\author{A. Lambrecht}
\affiliation{Laboratoire Kastler-Brossel ENS, UPMC, CNRS, Case 74,
Campus Jussieu, 75252 Paris Cedex 05, France}

\begin{abstract}
In this paper we study the role of surface plasmon modes in the
Casimir effect. The Casimir energy can be written as a sum over the
modes  of a real cavity and one may identify two sorts of modes, two
evanescent surface plasmon modes and propagative modes. As one of
the surface plasmon modes becomes propagative for some choice of
parameters we adopt an adiabatic mode definition where we follow
this mode into the propagative sector and count it together with the
surface plasmon contribution, calling this contribution
``plasmonic". We evaluate analytically the contribution of the
plasmonic modes to the Casimir energy. Surprisingly we find that
this becomes repulsive for intermediate and large mirror
separations. The contribution of surface plasmons to the Casimir
energy plays a fundamental role not only at short but also at large
distances.  This suggests possibilities to taylor the Casimir force
via a manipulation of the surface plasmons properties.
\end{abstract}

\maketitle

\section{Introduction}


The Casimir force is the archetypal mechanical consequence
of vacuum fluctuations in the quantized electromagnetic field. 
In its simplest form, it gives rise to the attraction of two planar
mirrors placed in empty space at zero temperature \cite{cas}.
The corresponding interaction energy $E$ takes a universal form
for perfect reflectors,
\begin{equation}
E = E_{\text{Cas}}=-\frac{\hbar c A}{ 4\pi \aleph  L^{3}}
,
\label{ECas}
\end{equation}
where $L$ is the distance between the mirrors, $A$ their area, and
$\hbar$ and $c$ the reduced Planck constant and the speed of light.
We abbreviate $\aleph = 180/\pi^3 \approx 5.8052762$.  As usual in
thermodynamics, a negative energy corresponds to a binding energy.
%

The Casimir force was soon observed in different experiments which
confirmed its existence
\cite{Sparnaay:1989,milonni,LamoreauxAmJPhys99}.  In recent years,
technological improvement allowed to reach a precision in the percent
range, which makes an accurate comparison to theoretical predictions
possible and has prompted a series of refined calculations
\cite{most,milton}.
Casimir's 1948 derivation of Eq.(\ref{ECas}) is based on summing the
zero-point energies $\frac12 \hbar \omega$ of the cavity
eigenmodes, taking the difference for finite and infinite separation,
and removing the divergences by inserting a high-energy cutoff.
He considered an ideal setting with perfectly reflecting mirrors
in vacuum.  Experiments are however performed with real reflectors,
typically metallic mirrors which are good reflectors only at
frequencies below the plasma frequency ($\omega_{\rm p} / 2 \pi$) or
alternatively at wavelengths much larger than $\lambda_{\rm p} = 2\pi
c / \omega_{\rm p}$. It has been known since a long time
that this has a significant effect on the force, in particular
at mirror distances of the order of $\lambda_{\rm p}$ or smaller
\cite{Lifshitz56,Heinrichs75,Schwinger78}, and 
precise investigations have been developed recently
\cite{Lamoreaux99,LambrechtPRL00,Lambrecht00,Genet00,most,%
KlimPRA00,MostepanenkoPRA00,Genet02,Pirozhenko06,Mohideen06}.

A system made from real material mirrors sustains electromagnetic
modes which strongly differ with respect to the ideal case, in
particular plasma oscillations and surface plasmons (sometimes called
surface plasmon polaritons).  These are collective electron density
waves with energies $\hbar\omega_{\rm p}$ around ten electron volts
(at typical metallic densities).  These waves can be quantized and
since $\hbar\omega_{\rm p}$ is larger than any experimentally relevant
thermal energy, one can safely consider that bulk plasma modes are in
the ground state \cite{Raimes57}.  This is not quite true for the
surface plasmon modes that are confined to the surface of a metallic
mirror.  Their electronic excitation is accompanied by an
electromagnetic field mode that is evanescent inside the cavity
\cite{Raether}.  Surface plasmons play an important role in many
fields of physics.  Let us only mention the plasmon-assisted light
transmission through metallic structures
\cite{ebbesen:nature,ebbesen:theory,woerdman}, or dispersion forces
between electronic Wigner crystals that are relevant for biomolecular
physics~\cite{Lau:2000}.  More generally, evanescent electromagnetic
waves have a strong impact on the Casimir-Polder interaction between
an atom and a surface as well as on the interaction between two
surfaces at differences temperatures~\cite{Obrecht07a,Antezza06b}.

It is well known, indeed, that the Casimir effect, at short distances,
is dominated by the coupling between the surface plasmons that
propagate on two metallic mirrors.  This has been pointed out in 1968
by Van Kampen and co-workers \cite{vankamp} who computed the Casimir
energy for $L \ll \lambda_{\rm p}$ in terms of quasi-electrostatic (or
\emph{non-retarded}) field modes.  In this limit the Casimir energy
becomes~\cite{Lifshitz56,Lambrecht00,genet:vacuum}
\begin{equation}
\label{shortapprox} E\approx \alpha\frac{L}{\lambda_{\rm p}}
E_{\text{Cas}}
\quad\text{with
$\alpha \approx 1.790$}
.
\end{equation}
which is smaller than Eq.(\ref{ECas}).  Observe the different power
law and the non-universal behavior as the result depends on the
material parameter $\lambda_{\rm p}$.  For metals used in modern
experiments, $\lambda_{\rm p}$ lies in the sub-micron range (107nm for
Al and 137nm for Cu and Au).  This short-distance regime has been 
studied in much detail since Van Kampen's paper, investigating, for 
example, materials with a nonlocal response~\cite{Heinrichs75a,Summerside79}.


As the mirror separation increases, \emph{retardation} has to be taken into 
account, and Van Kampen's result calls for a generalization.
This has been done by Schram in 1973
\cite{schram}, improving on a previous paper by Gerlach 
\cite{gerlac}.
Schram considered mirrors described by a
non-dissipative dielectric function, found
the electromagnetic modes vibrating between these mirrors, and 
got the Casimir energy by summing their zero point energies.
Among these modes, we find the retarded version of van Kampen's surface
plasmon modes. Schram did not analyze separately their 
contribution and focused on the total energy, using a calculation
based on the argument principle.  
Summerside and Mahanty investigated
the joint effect of retardation and nonlocality on the surface plasmon
modes at short distances \cite{Summerside79}.

In this paper we investigate more closely the influence of
surface plasmon modes on the Casimir energy, covering both the 
non-retarded and retarded domains. 
This permits to explore the experimentally relevant distance range
around one micron where current precision experiments are performed.
The plasmon modes are identified in a natural way in the sum over
electromagnetic modes of the real cavity. 
We have shown previously that they have peculiar properties  
\cite{intravaia:110404}:
one of them is purely evanescent while the dispersion relation of
the other one changes its character from
evanescent to propagating inside the cavity (it crosses the light cone).
In addition, the combined plasmonic contribution
to the Casimir energy has the peculiarity to change sign as a
function of distance $L$.
Here, we derive and expand on these results in more detail and
exhibit closed-form expressions valid at all distances.
The main idea is to 
perform a re-parametrization of the dispersion relations that 
permit to evaluate analytically the relevant integrals.
We recover van Kampen's result at short distances and discuss
explicitly the asymptotic behaviour 
in the long distance domain where 
retardation plays an important role.
This regime was not covered in a previous paper by one of us
\cite{henkel:023808} that performs an analysis of surface plasmons
in the short-distance (non-retarded) regime.
The analysis of the ``photonic modes'' (corresponding to waves
that propagate in the cavity) will be the object of a following 
paper. For simplicity, we restrict here to zero temperature, 
the generalization to finite temperature being straightforward.

The material is organized as follows. The basic method and the 
cavity modes are introduced in Sec.~II. The dispersion relation
of the plasmonic modes is analyzed in Sec.~III and
Appendix~\ref{s:dispersion-relations}, and their contribution
to the Casimir energy given in Sec.~III\,A. The Secs.~III\,B and~C
discuss the short and large distance regimes. Our analysis concludes
with a discussion of the sign of the Casimir interaction (Sec.~III\,D)
and of alternative splittings of the plasmonic 
dispersion relations that appeared recently in the literature (Sec.~III\,E).

\section{Casimir interaction and real cavity modes}
\label{s:real-cavity-modes}

In 1973 Schram proved the following mathematical identity
\cite{schram}, exploiting the argument principle \cite{
most}
\begin{widetext}
\begin{equation}
\label{en}
E =
\sum_{\mu,\mathbf{k}}\left[\sum_n \frac{\hbar}{2}\omega^{\mu}_n( {\bf k} )
\right]^L_{L\rightarrow\infty}
= {\rm Im}\, \sum_{\mu,\mathbf{k}}\int\limits_{0}^{\infty }
\frac{\mathrm{d}\omega }{2\pi}
\hbar
\ln({1-r_{\mathbf{k}}^{\mu}[\omega]^{2}e^{2ik_{z}L}})
\end{equation}
\end{widetext}
The left-hand side has the same structure as 
Casimir's sum over zero point energies, 
but in this case the relevant modes are those of the real cavity. 
The notation $\left[  \cdots \right]^{L}_{L \to \infty}$ signifies 
the difference of the expression in brackets for finite and infinite 
mirror distance $L$.  The right-hand side is nothing but the Lifshitz
formula for the Casimir energy. Let us recall that Lifshitz
adopted in 1955 \cite{Lifshitz56} a fairly different viewpoint
and computed the force as the average of the Maxwell stress tensor
inside the cavity. He considered the electromagnetic fields as being
radiated by fluctuating sources in the medium composing the mirrors,
similar to London's derivation of the Van der Waals force between atoms
and molecules. The main point of Ref.\cite{schram} was to show that 
the Lifshitz approach yields the same result as the Casimir sum over 
zero-point energies, provided the mirrors are non-dissipative. This 
is the case we focus on here.

The modes in Eq.(\ref{en}) are labelled by their
polarization $\mu={}$TE, TM and the wavevector
$\mathbf{k}\equiv \left( k_{x},k_{y}\right)$ parallel to the
mirrors; the perpendicular wavevector $k_z$ is defined in  
Eq.\eqref{defkappa} below.
The $r_{\mathbf{k}}^{\mu}$ are the reflection amplitudes that we take
the same for both mirrors. The mode frequencies $\omega^{\mu}_n( {\bf k} )$ 
are related to the zeros and the branch cuts of~\cite{schram}
\begin{equation}
\label{modes1} 
D_{\mu}[ \omega; {\bf k} ] 
= 1 - r_{\mathbf{k}}^{\mu}[ \omega ]^{2}e^{2ik_{z}L}
.
\end{equation}
We adopt here the Fresnel formulas for the reflection amplitudes that
for the case of thick mirrors read \cite{bornwolf}
\begin{equation}
\label{tetemrefdef}
r^{TE}=\frac{\kappa -\kappa_{\rm m}}{\kappa +\kappa_{\rm m}},\quad
r^{TM}=\frac{\kappa_{\rm m}-\epsilon[\omega]\kappa 
}{\kappa_{\rm m}+\epsilon[\omega]\kappa}
\end{equation}
where 
\begin{subequations}
\label{defkappa}
\begin{equation}
k_z = \imath \kappa = \imath
\sqrt{\abs{\mathbf{k}}^{2}-\omega ^{2} / c^2}
\end{equation}
\begin{equation}
\label{defkappam} 
\kappa_{\rm m}=\sqrt{\abs{\mathbf{k}}^{2}-\epsilon[\omega]\omega
^{2} / c^2 }=\sqrt{\kappa^2 + \omega _{\rm p}^{2} / c^2}
\end{equation}
\end{subequations}
We choose signs for the square roots such that
$\re{\kappa_i}>0$ and $\im{\kappa_i}<0$  in $\im\omega>0$.
This analytical continuation 
entails that Eq.(\ref{modes1}) has no solutions in the
upper half plane. 
Finally, 
$\epsilon[\omega]$ is the dielectric
function; in the case of a metal the simplest description is given
by the plasma model
\begin{equation}
\epsilon[\omega]=1-\frac{\omega_{\rm p}^{2}}{\omega^{2}}
\end{equation}
where $\omega _{\rm p}$ is the plasma frequency, a constant which can be 
related 
to the specific physical properties of the metal. Up to
$\omega\sim\omega_{\rm p}$ the dielectric constant differs from unity 
so that the metal behaves different than the surrounding 
vacuum. For $\omega\gg\omega_{\rm p}$ the dielectric constant approaches 
unity and the metal becomes transparent. This is the way the 
plasma model implements the high-frequency cutoff for the mirror 
reflectivity.

%
\begin{figure}[bth]
   \centerline{%
\includegraphics*[width=7.5cm]{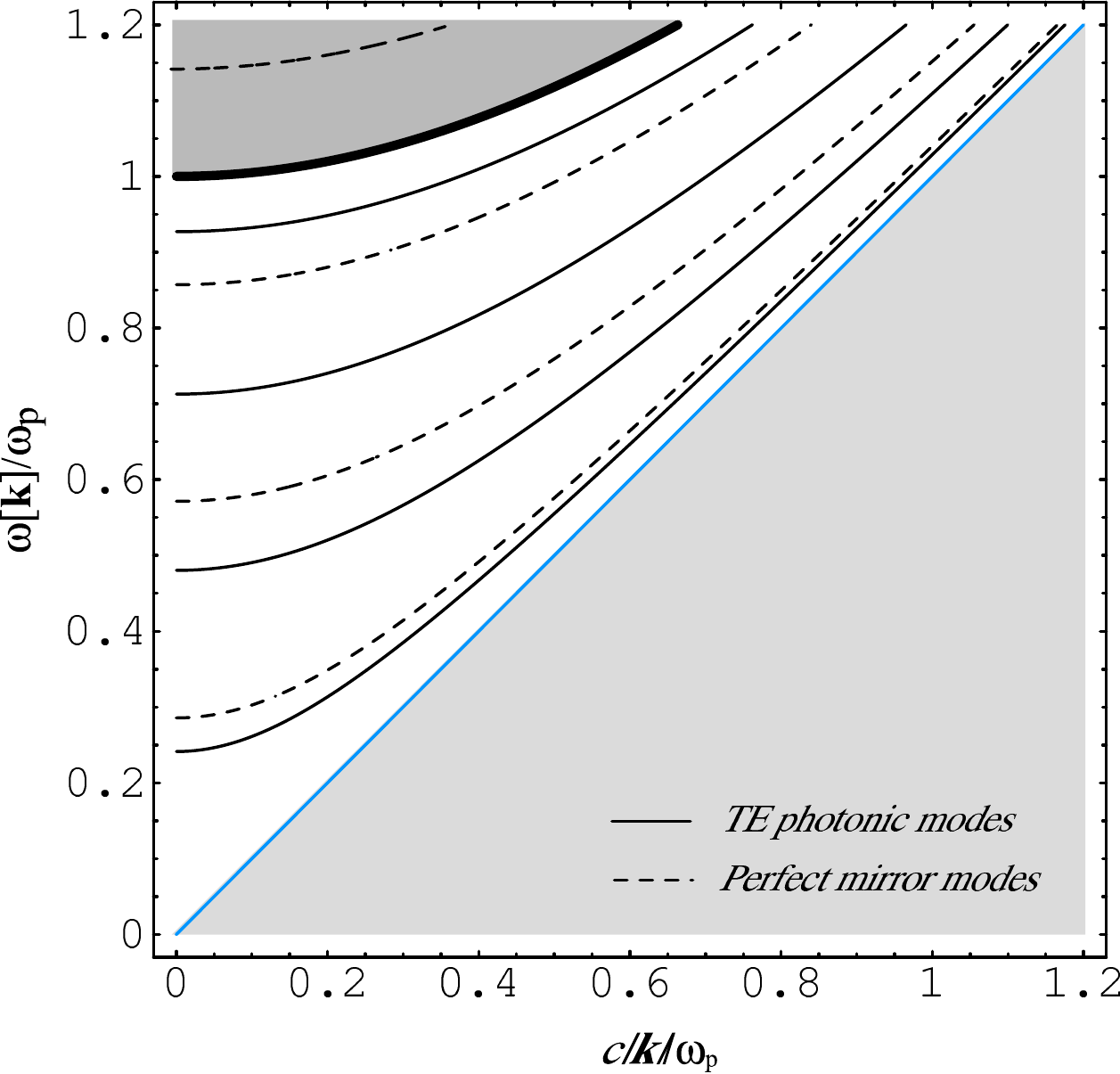}
}
   \caption{Dispersion relations for TE-polarized modes 
between two metallic
mirrors described by the plasma model (solid line), compared
perfect conductors (dashed line). Mode frequency $\omega( {\bf k} )$
and wavevector in the mirror plane, $|{\bf k}|$, are normalized 
to the plasma frequency $\omega_{\rm p}$. Mirror distance
$L = 1.75\,\lambda_{\rm p}$. The (blue) diagonal line is the light cone
below which the field is evanescent in the cavity (evanescent modes). 
Above the thick solid line, the field propagates through in mirror
material (bulk modes). 
}
   \label{dispte}
\end{figure}

In this model we neglect all the dissipation phenomena and we impose
a local response to the electromagnetic field  \cite{bornwolf}. 
From a physical point of view, it is a poor approximation to real
metals at low frequencies (dissipation and non-locality, i.e., the
anomalous skin effect are predominant) and high frequencies 
(absorption
from intraband transitions). But at any rate, its
mathematical simplicity allows explicit calculations to be pushed
very far and to understand important physical behaviors. We are going 
to see
that our principal result correspond to a frequency range high
enough for the plasma model to be a good description of the metal. 
Let us stress, however, that the choice of the plasma model is the 
strongest approximation we make and, within this model, all results 
we discuss are exact.

\begin{figure}[bth]
   \centerline{%
\includegraphics*[width=7.5cm]{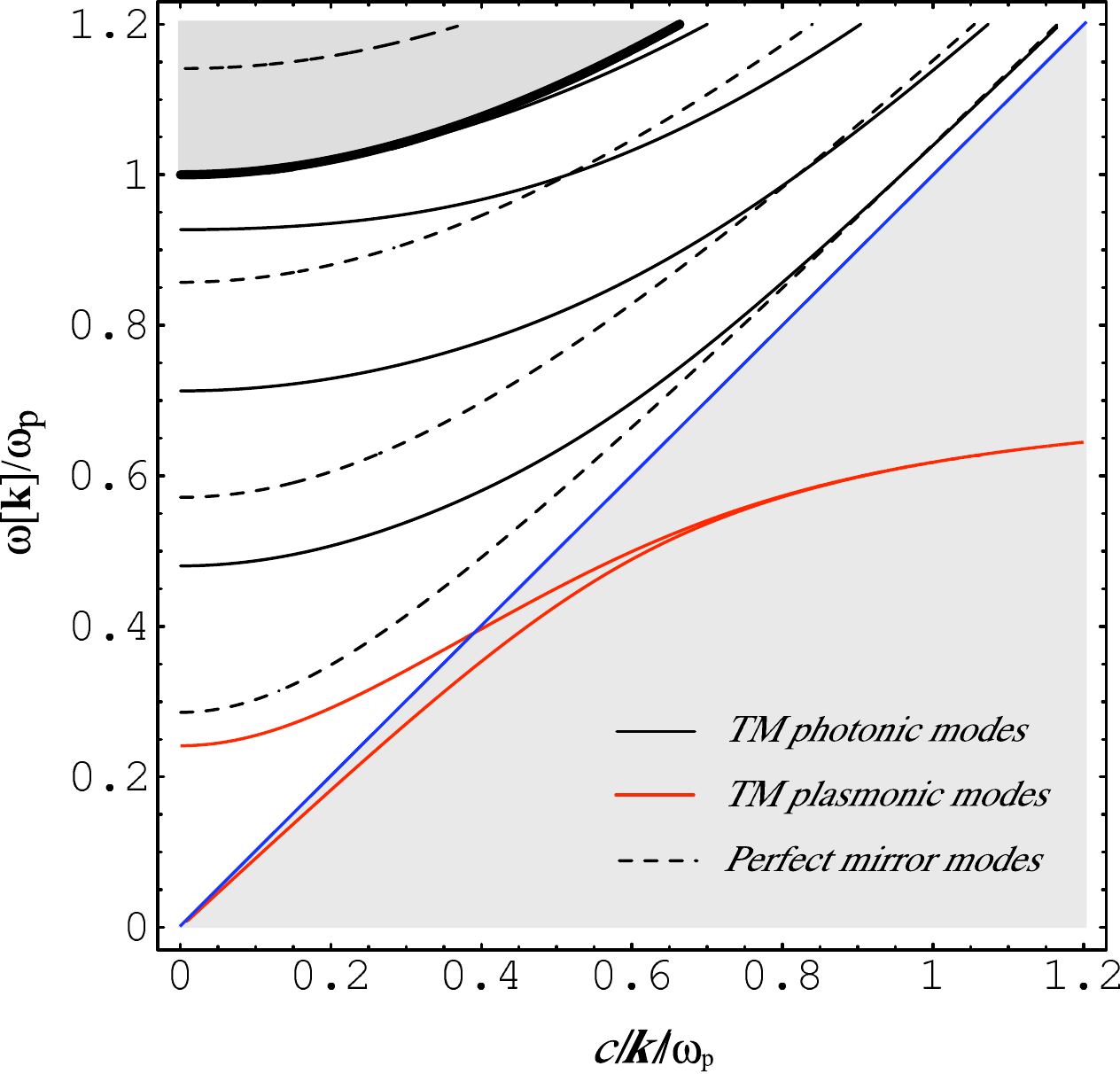}
}
\caption{Same as Fig.\ref{dispte} for TM-polarized modes.
The solid 
red curves represent the plasmonic modes and the black curves the
   photonic modes. Note that one of the plasmonic modes crosses
the light cone.}
   \label{disptm}
\end{figure}
%

The introduction of the dielectric properties of the
mirrors leads to a series of important modifications for the
field modes. First of all,
even in the simplest case, the plasma model, the dispersion relations
$\omega^{\mu}_{n}( {\bf k} )$ cannot be written in terms of
elementary functions.  The results of a numerical calculation are
shown in figs. \ref{dispte} and \ref{disptm} (see details below).
As we can see, imperfect reflection modifies the dispersion relation
(solid lines) compared to a perfect reflector (dashed lines).

We can distinguish three regions
starting from above:

\textit{Bulk modes} occur for $\omega > \omega_B( {\bf k} )
= (\omega^{2}_{\rm p} + c^2
|\mathbf{k}|^{2})^{1/2}$ (shaded above the thick line); they propagate
both in the cavity and inside the mirrors.  These modes form a 
continuum that is mathematically
represented by a branch cut of Eq.(\ref{modes1})
in the complex 
$\omega$-plane. This has to be taken into account carefully when applying
the argument theorem \cite{marku}. The associated difficulties
have led Schram to work instead with a mirror of finite thickness $d$
where the continuum discretizes \cite{schram}. For simplicity, we take 
here the limit of thick mirrors.


\textit{Propagating (ordinary) cavity modes}: they occur in the region
above the light cone and below the bulk continuum, $c\abs{\mathbf{k}}
< \omega < \omega_{B}( {\bf k} )$.  These modes are guided between the 
mirrors (note that the latter behave like a medium 
optically thinner than vacuum, $0 < \epsilon[ \omega ] < 1$),
leading to a discrete set of mode frequencies for a given ${\bf k}$.
In this region, the reflection coefficients~(\ref{tetemrefdef})
have unit modulus and a frequency-dependent phase. This 
leads to a shift of the cavity modes relative to perfectly 
reflecting mirrors, as is visible in Figs.\ref{dispte},\ref{disptm}.

\begin{figure}[tbh]
    \centerline{%
    \includegraphics*[width=6.5cm]{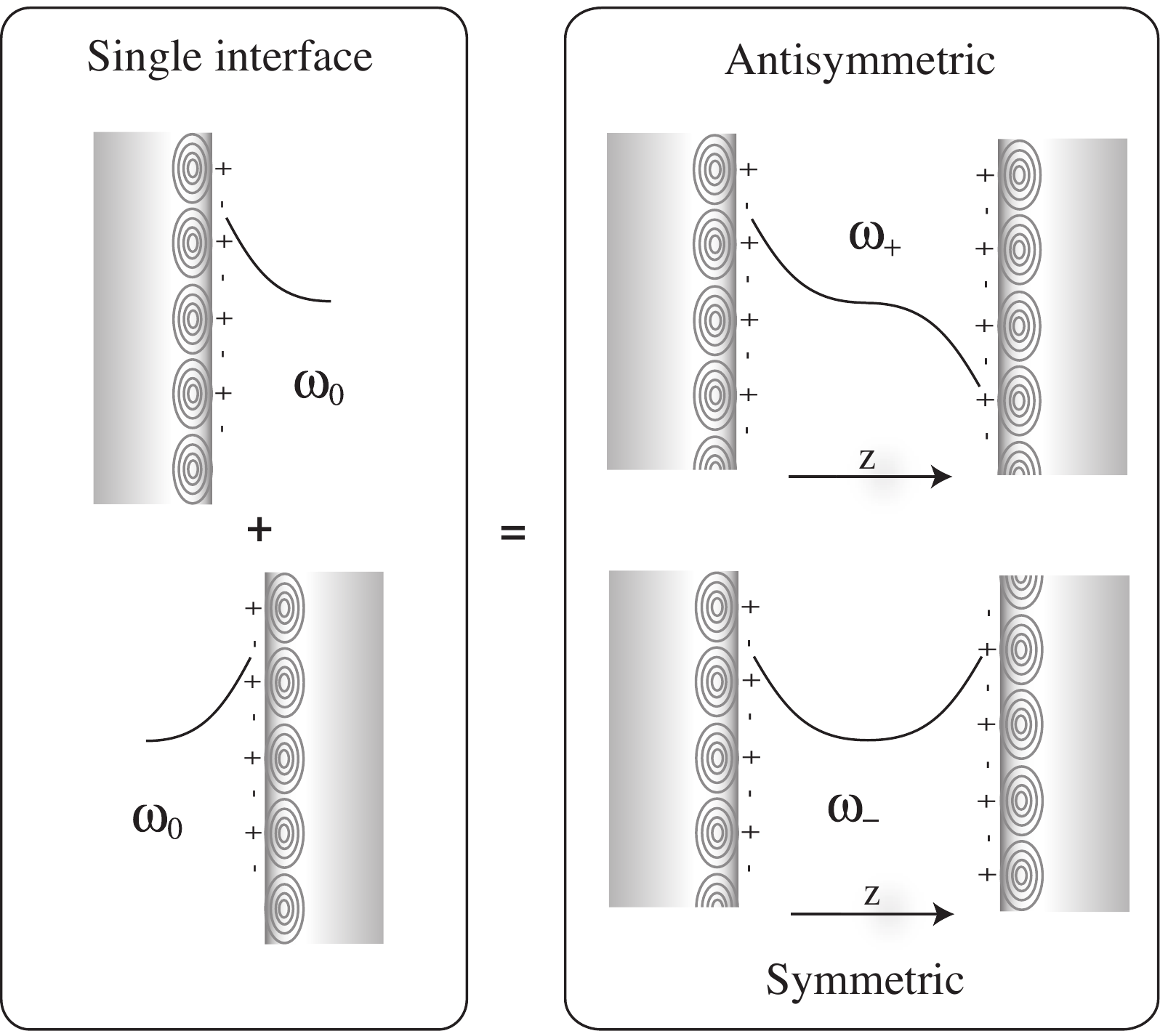}
    }
\caption{%
A surface plasmon mode is associate with an electronic charge
oscillation bound to the surface of a body.  For a single body the
associated electric field is evanescent and, for a plane interface,
the plasmon can be excited only by approaching from the vacuum side
a medium with higher index of refraction and illuminating the latter
in total internal reflection.
Approaching two surfaces, the two respective surface plasmons
couple through their evanescent field tails.  A frequency splitting
occurs giving rise to two new modes, the plasmonic modes.  The 
antisymmetric ($\omega_{+}$) and the symmetric ($\omega_{-}$) mode have
higher resp.\ lower energy than the isolated (non-coupled)
mode $\omega_{0}$.  The Casimir force associated with $\omega_{+}$
is then an anti-binding force (repulsive) while the $\omega_{-}$
modes contribute an attractive force.  The plasmonic Casimir force
arises from the (distance-dependent) balance of the two contributions.}
  \label{pls}
\end{figure}

\textit{Evanescent modes} lie below the light cone,
$\omega < c \abs{\mathbf{k}}$  (shaded below the diagonal),
and are the main focus of this paper.
Their electromagnetic field 
exponentially decreases when going away from the vacuum-mirror
interface, while it is allowed to propagate along the interface.
Evanescent fields are of great interest in near field optics because
they provide the link to sub-wavelength topographic features of a 
surface. In the context of the Casimir interaction, they
are often underestimated due to their damped nature. We show here,
however, that their contribution is all but a small correction,
even at large distances \cite{intravaia:110404}.  From a mathematical
point of view, the
optical properties of evanescent modes (reflection and transmission
amplitudes) can be obtained from ordinary modes by a well-defined
analytical continuation procedure \cite{Genet03c}. Solving the
dispersion equation~(\ref{modes1}) in the evanescent sector, one
finds two nondegenerate mode frequencies in only one polarization,
at least for non-magnetic media.
These modes 
are called ``surface plasmons'' (or ``surface plasmon
polaritons'')~\cite{Raether,Economou69,Chang73}.
Their field amplitude 
decays exponentially away from the interface, 
and is associated with oscillating surface charge and surface current
densities, as required by the equation of continuity (see
Fig.\ref{pls}).  
On an isolated interface, surface plasmons correspond to the pole of
$r^{TM}_{\bf k}[ \omega ]$; they occur when $\varepsilon[ \omega ] <
-1$ (i.e., $\omega < \omega_{\rm p} / \sqrt{ 2 }$).
For two interfaces, two surface plasmons exist and are 
coupled via their evanescent tails in the cavity.  The resulting
modes are given
by the zeros of Eq.(\ref{modes1}) for real $\kappa$ and $\kappa_{\rm m}$,
and will be analyzed in detail in the following. 

Summarizing, 
we can see that for the TE-polarization, all modes lie above
the light cone, while for TM-polarization, two modes enter the
evanescent region in at least some range of wavevectors. We refer
to these modes as ``plasmonic''; they are the retarded generalization
of van Kampen's coupled surface plasmon modes. Finally, we
can re-write the Casimir energy as
\begin{widetext}
\begin{equation}
\label{start} 
E=\underbrace{ 
\sum_{\mathbf{k}}\left[\frac{\hbar\omega_+}{2}
+\frac{\hbar\omega_-}{2}\right]^L_{L\rightarrow\infty}}_{\text{plasmonic
modes ($E_{\mathrm{pl}}$)}}
+\underbrace{
\overbrace{\sum_{\mu,\mathbf{k}}\left[\sum_{\omega<\omega_B}
\frac{\hbar\omega^{\mu}_n}{2}\right]^L_{L\rightarrow\infty}}^{\text{cavity modes}}
+
\overbrace{
\lim\limits_{d\to\infty}\sum_{\mu,\mathbf{k}}\left[\sum_{\omega\ge\omega_B}
\frac{\hbar\omega^{\mu}_n}{2}\right]^{L,d}_{L\rightarrow\infty,d}}^{\text{bulk
modes}}}_{\text{photonic modes
($E_{\mathrm{ph}}$)}}
\end{equation}
\end{widetext}
These contributions have no physical meaning on their own,
i.e. one cannot measure them separately. 
The only observable  is the
total Casimir energy, which is the sum of all terms.
However, evaluating them separately reveals striking features
which suggest new possibilities to taylor the strength and the sign
of the Casimir force.  In the rest of this paper, we are going to focus our
attention on the plasmonic contribution $E_{\rm pl}$ and
shall discuss the remaining contributions to the Casimir energy in
another paper. 

\section{Plasmonic Modes}

\begin{figure}[tbh]
    \centerline{%
 \includegraphics*[width=8cm]{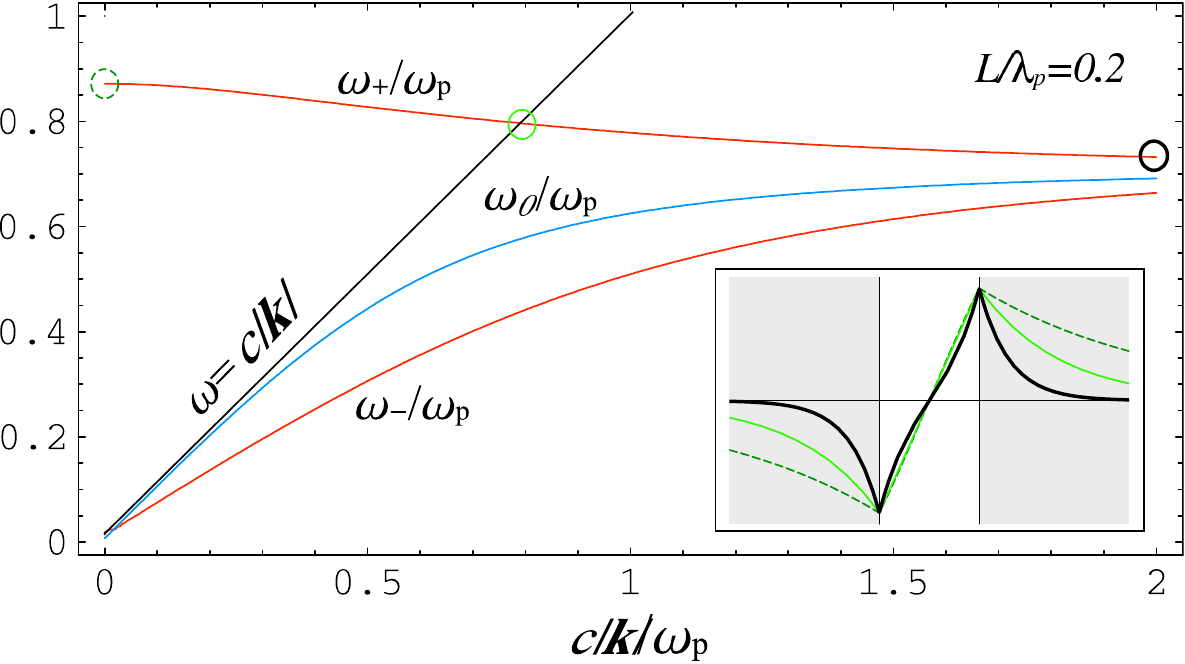}
 }
\caption{A plot of the plasmonic dispersion relations
  $\omega_{+}( {\bf k} )$,
  $\omega_{0}( {\bf k} )$,
  $\omega_{-}( {\bf k} )$, as function of 
$\abs{\mathbf{k}}$  for
  $L = 0.2\,\lambda_{\rm p}$ ($\lambda_{\rm p} = 2\pi c / \omega_{\rm p}$).
  Frequencies and wavevectors are scaled to the plasma frequency
  $\omega_{\rm p}$ and $\omega_{\rm p}/c$, respectively. 
  Inset: magnetic field amplitude for chosen points along the 
  branch $\omega_{+}( {\bf k} )$, as labelled by the circles.}
  \label{opm}
\end{figure}

We plot again in Fig.\ref{opm} the dispersion relation of the
two modes in the first sum of Eq.(\ref{start}). They end up 
for large $|{\bf k}|$ below the light cone, i.e., the associated
field is evanescent both in vacuum and in the mirrors. One branch that 
we call $\omega_{-}( {\bf k} )$ lies entirely below the light cone.
The second one, $\omega_{+}( {\bf k} )$ moves continuously into the
cavity mode sector as ${\bf k}$ is decreased. The inset illustrates
the smooth change in the spatial mode function. This mixed 
character justifies the name ``plasmonic'' that we use for both 
modes in the following. We discuss in Appendix~\ref{s:dispersion-relations}
some general 
features of their dispersion relations that can be obtained explicitly 
despite the fact that we have to deal with implicit functions. 

\subsection{Contribution to the Casimir energy}
\label{s:plasmonic-Casimir}

The plasmonic contribution is defined as the first sum on the r.h.s. 
of eq.\eqref{start}, namely
\begin{equation}
\label{contr1}
E_{\rm pl}=\sum_{\mathbf{k}}\left[\frac{\hbar\omega_+}{2}
+\frac{\hbar\omega_-}{2}\right]^L_{L\rightarrow\infty}
\end{equation}
Both modes tend to 
$\omega_{0}( K )$ for $L \to \infty$ so that we subtract the 
zero-point energy for two isolated surface plasmons. We are thus
measuring the interaction energy arising from the coupling between 
the surface plasmons.

Replacing the ${\bf k}$-summation by an integral and using the scaled 
variables introduced in~(\ref{eq:scaled-variables}), 
we get
\begin{equation}
\label{contr}
E_{\rm pl} = \frac{ \hbar c A }{ 2 L^3 } 
\int\limits_{0}^{\infty}\frac{K dK}{2\pi}
\left(
\Omega_+( K )
+
\Omega_-( K ) - 2 \Omega_{0}( K ) 
\right).
\end{equation}
To check the convergence at large $K$, we use the 
parametrization of Eqs.(\ref{eq:z-parametrization})
and find the estimate
\begin{equation}
    \Omega_\pm^2( K ) - \Omega_{0}^2( K ) \to \pm \frac12
    \Omega_{\rm p}^2 e^{- K } + O( e^{- 2K } )
    ,
    \label{eq:large-K-asymptotics-0}
\end{equation}
provided $K \gg \max(1, \Omega_{\rm p})$.
For further details, see the discussion around Eq.(\ref{converg}).
The difficulty in Eq.(\ref{contr}) is that the dispersion relations
$\Omega_{\pm}( K )$ are only known implicitly in the general case.  We
now show that using the parametrization of
Appendix~\ref{s:dispersion-relations}, the integrand can be brought
into an explicit and elementary form.

It is useful to scale the energy Eq.\eqref{contr} to the
perfect-mirror Casimir energy $E_{\rm Cas}$ [Eq.(\ref{ECas})]
\label{defplascontr}
\begin{equation}
E_{\rm pl} = \eta_{\rm pl} E_{\rm Cas}
\end{equation}
\begin{align}
\label{etam} \eta_{\rm pl} = -
\prefactor
\int_0^{\infty} \sum_{ a = \pm, 0} 
c_a \Omega_a( K ) \, K dK
\end{align}
with $c_+=c_-=1,c_0=-2$. We call $\eta_{\rm pl}$ the correction factor 
for the plasmonic Casimir energy; note that it depends on the distance
only via the dimensionless parameter $\Omega_{\rm p}$.

%

For each of the branches $\Omega_{a}( K )$, we now 
change to the integration variable $z = (\kappa L)^{2}$.
The Jacobian (the prime denotes the derivative)
\begin{equation}
\label{change}
d K^2 = 2 K dK = dz + g_a'(z) dz
\end{equation}
with $g_a(z)$ defined in Appendix~\ref{s:dispersion-relations}, 
leads to
\begin{equation}
    \eta_{\rm pl} = -
    \halfprefactor
    \sum_{ a = \pm, 0} c_a 
    \int\limits_{\Gamma_{a}} ( 1 + g'_{a}( z ) )
    \sqrt{ g_{a}( z ) } \, dz
    \label{eq:re-parametrize-integral}
\end{equation}
The integration paths are now $\Gamma_{+} = - z_{+} \ldots \infty$,
and
$\Gamma_{-,0} = 0 \ldots \infty$ where $z_{+}$ is defined in 
Eq.(\ref{eqy}).


One of the two terms under the integral can be integrated 
immediately, leading to
\begin{equation}
    \sum_{ a = \pm, 0} c_a 
    \int\limits_{\Gamma_{a}} g'_{a}( z ) 
    \sqrt{ g_{a}( z ) } \, dz
    =
    \frac23 
    \sum_{ a = \pm, 0} c_a 
    \left[ g_{a}^{3/2}( z ) \right]_{\Gamma_{a}}
    \label{eq:re-parametrize-integral-2}
\end{equation}
where the function in brackets has to be evaluated at the end points 
of the respective integration domains. The upper limit 
contributions ($z = \infty$) cancel under the subtractions.  At the
lower
limit, $g_{-,0}^{3/2}( 0 )$ vanishes because the dispersion relations 
reach $\omega = 0$ (see Sec.\ref{s:dispersion-relations}).
We are thus left with $g_{+}^{3/2}( -z_{+} ) = z_{+}^{3/2}$.

%
%

Putting the propagating sector of the mode $\Omega_{+}( K )$ into a 
separate integral, 
the correction factor for the plasmonic Casimir energy can be rewritten as
\begin{equation}
\label{fin}  
\eta_{\rm pl} = -\halfprefactor
\left[ 
\int_0^{\infty}\sum_a c_a
\sqrt{ g_a(z) }dz + 
\int_{-z_+}^{0} \sqrt{ g_+(z) }dz - \frac{2}{3}z_{+}^{3/2}
\right]
\end{equation}
In the first integral, the functions $g_a(z)$ are real. For 
$z \to \infty$, the functions $g_{\pm}( z )$ approach $g_{0}( z )$
exponentially fast.
An expansion in $e^{- \sqrt{z} }$ leads to
\begin{equation}
\label{converg} \sum_a c_a
\sqrt{ g_{a}(z) } \approx 
-\Omega_{\rm p} e^{-2\sqrt{z}} f( z / \Omega_{\rm p}^2 )
\end{equation}
where the function $f( z / \Omega_{\rm p}^2 )$ is bounded and tends to
$1/(4\sqrt{2})$ for $z \gg \Omega_{\rm p}^2$. This secures the convergence 
at large $z$ of the first integral in Eq.(\ref{fin}).
The second integral is finite because $g_{+}( z )$ is
bounded and the integration domain is finite [see
Eq.(\ref{eq:zplus-limits})].  
Both the second integral and the third term in Eq.(\ref{fin})
are related to the propagating segment of the 
plasmonic mode $\Omega_{+}( K )$.

The great advantage of Eq.\eqref{fin}
compared to Eq.\eqref{contr} is that now the integrands are expressed in
terms of simple analytic functions and there is no need to integrate
implicit functions whose evaluation is only possible numerically.
We also gain for analytical calculations 
since the discussion of the distance dependence (via the parameter 
$\Omega_{\rm p} \propto L / \lambda_{\rm p}$) can be done in a transparent way.
%
We show in the following that one gets
asymptotic expressions
for small and large values of $\Omega_{\rm p}$, the only variable
on which the correction factor $\eta_{\rm pl}$ depends after the
integration. 

Fig.\ref{pat} shows a plot of $\eta_{\rm pl}$ as function of $L/
\lambda_{\rm p} = \Omega_{\rm p} / (2\pi)$.  Note the increase linear
in $L$ for small distances and a sign change at large $L$, with a
power law $\propto L^{1/2}$.  In the next two sections we analyze
these limits analytically.  At short distance $\eta_{\rm pl}$
reproduces exactly the correction factor known for the total Casimir
energy.

\subsection{Short distance asymptotics}
\label{s:asymptotics}


The distance enters the correction factor $\eta_{\rm pl}$ 
[Eq.(\ref{fin})] via the dimensionless parameter $\Omega_{\rm p}$,
and we get the short-distance asymptotics in the limit 
$\Omega_{\rm p}\ll1$. This has been discussed in previous papers
\cite{genet:vacuum,gerlac,henkel:023808,Bordag:2006}, but the
asymptotics turns out to be tricky at next-to-leading order.

The first order expansion in $\Omega_{\rm p}$ of the functions $g_{a}( z )$ 
yields~\cite{genet:vacuum}
\begin{subequations}
    \begin{equation}
    \eta_{\rm pl} \approx \alpha \frac{ \Omega_{\rm p} }{ 2 \pi }
    = \alpha \frac{ L }{ \lambda_{\rm p} }
    \label{s}
\end{equation}
where the numerical constant $\alpha \approx 1.790$ arises from
\begin{equation}
    \alpha = - \frac{ \pi\, \prefactor }{ \sqrt{ 2 } }
    \int\limits_{0}^{\infty}\left(
    \sqrt{ 1 + e^{ - \sqrt{z} } } +     
    \sqrt{ 1 - e^{ - \sqrt{z} } } - 2
    \right) dz 
    \label{eq:def-alpha}
\end{equation}
\end{subequations}
The separate contributions of the modes $\Omega_{+}(K)$ and 
$\Omega_{-}(K)$ are $\alpha_{+} \approx - 12.225$ (repulsive)
and $\alpha_{-} \approx 14.015$ (attractive).
The plasmonic 
Casimir energy in this regime thus scales like $A \hbar \omega_{\rm p} / L^2$
and is reduced compared to the perfect mirror case ($\eta_{\rm pl} 
\ll 1$) \cite{Lambrecht97}.

The contribution of the propagating part of $\omega_+$ is of the 
third order 
in $\Omega_{\rm p}$ [see Eq.(\ref{eq:zplus-limits})]
\begin{equation}
\int_{-z_+}^{0}g_+(z)dz 
\xrightarrow{\Omega_{\rm p}\ll1}
g_+(0)z_{+} \approx \Omega_{+0}\Omega_{\rm p}^2 \approx \Omega_{\rm p}^3
\end{equation}
and can therefore be neglected.  The same argument holds for the term
$(z_+)^{3/2}\approx \Omega_{\rm p}^3$.  
In other words, the plasmonic contribution comes
essentially from the evanescent sector ($z>0$).
We note that the result~(\ref{s}) yields exactly the short-distance 
behavior of the \emph{full} Casimir 
energy
which is thus dominated at short distance by the interactions between 
surface plasmons~\cite{genet:vacuum,gerlac,henkel:023808,Bordag:2006}.

\begin{figure}[tbh]
    \centerline{%
 \includegraphics*[width=8.5cm]{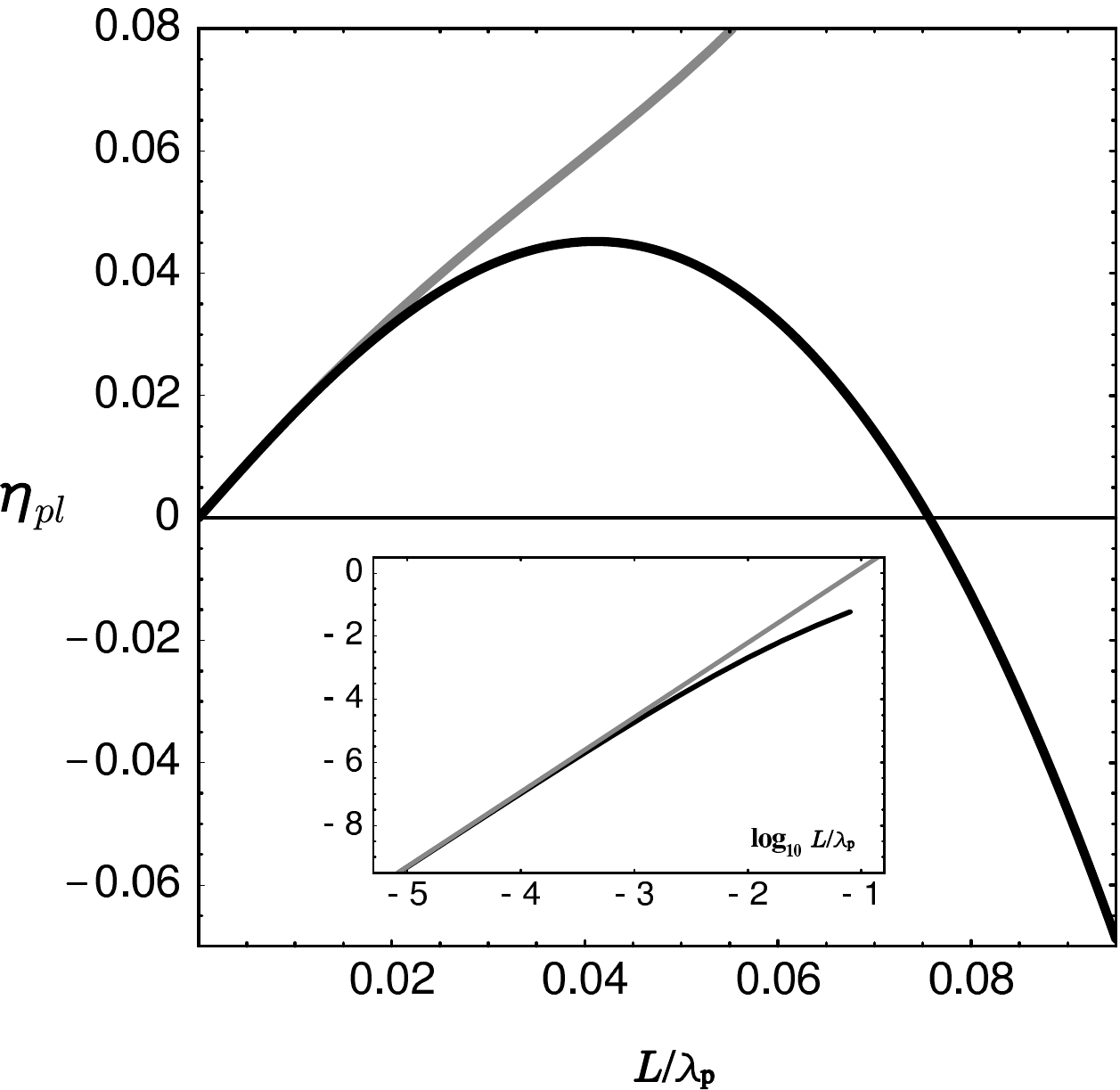}
 }
\caption{%
A plot of the plasmonic Casimir energy, normalized to the perfect 
mirror case (i.e., the plasmonic correction factor $\eta_{\rm pl}$) 
vs.\ the scaled distance $L / \lambda_{\rm p}$. 
Light gray solid line:
approximation~(\ref{eq:eta-short-3rd}).
The inset illustrates the short-distance behaviour beyond the linear
term: 
we plot $[ \eta_{\rm pl} - \alpha (L / \lambda_{\rm p})] / \Omega_{p}^3$
(black line) and compare to the expression
$a + b \log (2\pi L / \lambda_{\rm p})$ in Eq.(\ref{eq:eta-short-3rd})
(light gray line).
}
  \label{pat}
\end{figure}


Note that Eq.\eqref{s} follows from an expansion of
the $g_a(z)$ to first order in $\Omega_{\rm p}$. It is worth
stressing that this expansion scheme does not work at higher orders,
the series being an asymptotic one and not uniformly convergent.
Each integral
obtained by this method at higher orders is divergent, 
except the first one given in
Eq.\eqref{s}. To avoid this problem, we use an alternative method
and write the functions $g_{\pm}( z)$ as follows
\begin{equation}
g_{\pm}(z) = g_{0}(z) \frac{1\pm e^{-\sqrt{z}}}{1 \pm \rho},
\end{equation}
with
\begin{equation}
    \rho = 
    e^{-\sqrt{z}} 
    \frac{ g_{0}( z ) - \Omega_{\rm p}^2 / 2}{ \Omega_{\rm p}^2 / 2 }
    \label{eq:def-rho}
\end{equation}
Now for $z > 0$, $|\rho|$ is bounded by unity and decays rapidly to 
zero as $z \to \infty$ \cite{genet:vacuum}. 
To compute the integral of $[g_{\pm}( z 
)]^{1/2}$, we expand in powers of $\rho$ and get the series
\begin{equation}
[ g_{\pm}(z) ]^{1/2} = \sqrt{ g_{0}(z)( 1\pm e^{-\sqrt{z}} ) } 
\sum\limits_{n=0}^{\infty}
\frac{\sqrt{\pi}}{\Gamma(\frac{1}{2}-n)}\frac{(\pm 
\rho)^{n}}{n!}
\end{equation}
where $\Gamma( \cdot )$ is the gamma function.
Taking the $n = 0$ term, the integration over $z$ leads 
to~\eqref{s}.
Higher order terms can be calculated explicitly, but the resulting
expressions are cumbersome and will not be reported here.
Including the next-to-leading order terms, we find
\begin{equation}
\eta_{\rm pl}\approx \alpha \frac{ \Omega_{\rm p} }{ 2\pi } 
+ (a + b\log \Omega_{\rm p})
\Omega^{3}_{\rm p}
\label{eq:eta-short-3rd}
\end{equation}
where $a \approx 0.63$ and $b = \prefactor / 4\sqrt{2} \approx
1.026$. This is plotted as gray line(s) in Fig.\ref{pat}, the
inset providing a zoom on the cubic and logarithmic terms 
(see caption).
Note that the term $a \Omega_{\rm p}^3$ gives a 
distance-independent correction to the Casimir energy and cancels 
when the force is computed. The presence of the logarithmic correction 
is due to the non-uniform convergence of the asymptotic series.
We find from~(\ref{eq:eta-short-3rd}) that the Casimir force does not 
feature a logarithmic correction at short distance.

\subsection{Large distance asymptotics}
\label{s:large-distance}

The curve $\eta_{\rm pl}( L )$ in Fig.\ref{pat} shows that the
plasmonic mode contribution is negative (repulsive)
at distances $L \gtrsim 0.08\,\lambda_{\rm p}$.
Mathematically, this can easily be 
seen from the large $\Omega_{\rm p}$ asymptotics of $\eta_{\rm pl}$.
One can check that the integrand of the first integral in
Eq.\eqref{fin} is
significantly different from zero only for $z\sim 1$. This 
suggests the following expansion of the $g_a( z )$ for 
$\Omega_{\rm p}\gg1$
\begin{subequations}
\begin{gather}
g_+( z ) \approx \sqrt{\Omega_{\rm p}}\sqrt{\sqrt{z}\coth(
{\textstyle\frac12}\sqrt{z} )}
\\
g_-( z ) \approx \sqrt{\Omega_{\rm p}}\sqrt{\sqrt{z}\tanh(
{\textstyle\frac12}\sqrt{z} )}
\\
g_0( z ) \approx \sqrt{\Omega_{\rm p}}\sqrt[4]{z}
\end{gather}
\end{subequations}
Moreover, the expansion to leading order in $\Omega_{\rm p}^2 \gg |z|$ can 
also be performed in the integral over the propagating sector
in~\eqref{fin} because the integration domain is limited to 
$-\pi^2 \approx z_{+} \le z \le 0$. Finally, we find that 
the integrated term in Eq.\eqref{fin} gives a negligible
contribution so that to leading order,
$\eta_{\rm pl} = - \Gamma \sqrt{ \Omega_{\rm p} }$~\cite{Bordag:2006}
with
\begin{eqnarray}
\label{gaexpre} \Gamma&=&\prefactor\int_{0}^{\infty}
y^{3/2}\left(\sqrt{\coth[\frac{y}{2}]} +
\sqrt{\tanh[\frac{y}{2}]}-2\right)dy \nonumber\\
&+&\prefactor\int^{\pi}_{0}y^{3/2}\sqrt{\cot[\frac{y}{2}]}dy
\end{eqnarray}
This expression can be evaluated numerically, giving as result
$\Gamma=29.75$
(i.e. the sum of 
$8.90$ ($+$ mode, evanescent sector),
$-7.23$ ($-$ mode, evanescent sector),
and
$28.09$ ($+$ mode, propagating sector). Note the large contribution of 
the propagating segment and the near cancellation of the two 
evanescent branches.

Since $\eta_{\rm pl}$ is negative at large distances, 
the plasmonic contribution provides a 
repulsive contribution to the Casimir interaction that scales like
$+ A \hbar \sqrt{ \omega_{\rm p} cÊ} / L^{5/2}$.
This is balanced 
in the \emph{total} Casimir energy by the contributions of photonic 
modes (cavity and bulk modes), recovering the attractive large-distance 
power law $E_{\rm Cas} \propto - A \hbar c / L^3$.


\subsection{Cancellations and signs}

We conclude our analysis by suggesting an interpretation of the signs 
of the plasmonic contributions to the Casimir energy. It is clear that
$E_{\rm pl}$ is due to the shift in the plasmon mode frequency relative
to the isolated interface [see Eq.(\ref{contr})]. 

\begin{figure}[hbt]
    \centerline{ \includegraphics*[width=75mm]{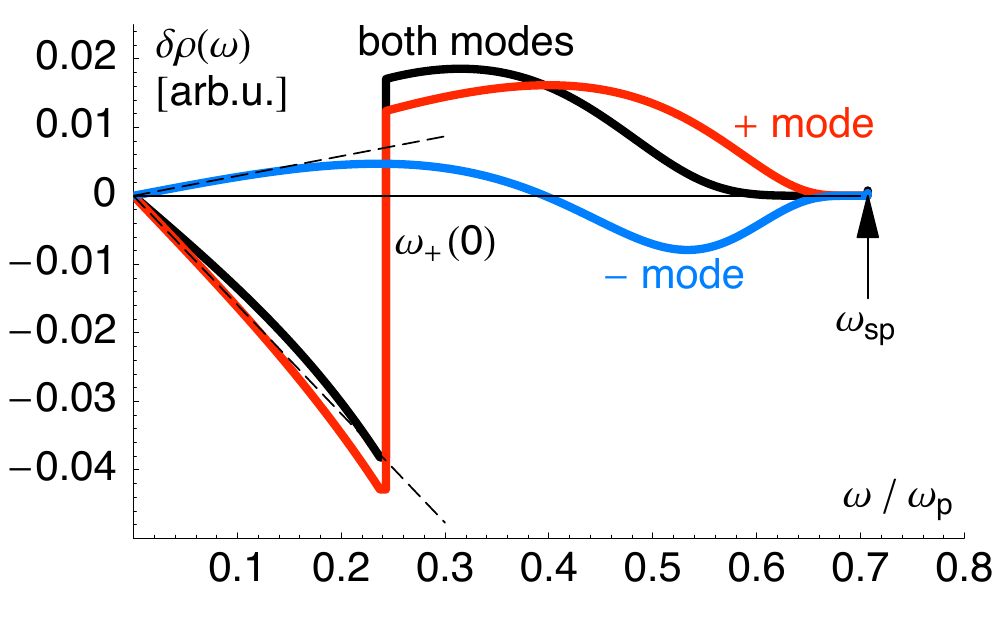}}
    \caption[]{Change in DOS, $\rho_{L,\pm}( \omega ) - 
    \rho_{\infty}( \omega )$, for the two plasmonic modes,
    as defined in Fig.\ref{opm}. 
    $L = 1.75\,\lambda_{\rm p}$ ($\Omega_{\rm p} \approx 10.996$).
    Dashed lines: small-frequency approximations discussed in the text.
    The frequency is scaled to $\omega_{\rm p}$, and the isolated surface 
    plasmon resonance is at $\omega_{\rm sp} = \omega_{\rm p}/\sqrt{2}$.}
    \label{fig:dos}
\end{figure}

This can be also interpreted as a reshuffling of the density of modes
due to the coupling by the interface, the total number of modes 
remaining constant. To make this more quantitative, we re-write the 
plasmonic Casimir energy as
\begin{equation}
    \frac{ \hbar }{ 2 }
    \left[\sum_{\textbf{k},a = \pm} \omega_{a}( \textbf{ k} )
    \right]_{L\rightarrow\infty}^{L}
    =
    \frac{ \hbar }{ 2 }
    \sum_{a = \pm}
    \int\limits_{0}^{\infty}\!{\rm d}\omega\,
    \omega
    \left[
    \rho_{a}( \omega )
    -
    \rho_{0}( \omega )
    \right]
    \label{eq:link-to-dos}
\end{equation}
where the mode densities are defined as usual by ($a = \pm, 0$)
\begin{equation}
    \rho_{a}( \omega ) =
    \sum_{\mathbf{k}} 
    \delta( \omega - \omega_{a}( \mathbf{k} ) ),
    \label{eq:def-dos}
\end{equation}
that depend on the distance $L$ for $a = \pm$.
Note that the $\omega$ integral
in~(\ref{eq:link-to-dos}) does not converge if taken over
the $\rho_{a}( \omega )$ alone. This is due to the flat large-$k$
asymptote of the plasmonic dispersion relations. More explicitly,
the density 
of modes can be calculated as
\begin{equation}
    \rho_{L,a}( \omega ) =
    \frac{ A \, k_{a}( \omega;L ) }{ 2\pi }
    \left| \frac{ {\rm d} k_{a}( \omega;L) }{ {\rm d} \omega }
    \right|
    \label{eq:dos-and-flatness}
\end{equation}
where, $k_{a}( \omega )$ is the inverse function to $\omega_{a}(
\mathbf{k} )$, and the derivative is just the inverse
group velocity at a given frequency $\omega$. We find
a behaviour $\rho_{a}( \omega ) \propto (\omega_{\rm sp}^2 - 
\omega^2)^{-2}$ when $\omega$ approaches the asymptotic value
$\omega_{\rm sp} \equiv \omega_{\rm p}/ \sqrt{2}$ of the dispersion relation 
(the surface plasmon resonance in the quasi-static limit).
This peak is exactly cancelled in the difference
$\delta\rho_{\pm}( \omega ) \equiv
\rho_{\pm}( \omega ) - \rho_{0}( \omega )$ that we plot
in Fig.\ref{fig:dos} for a given distance $L$. The precise behaviour 
of the curves changes with the distance (at smaller $L$, for example,
$\rho_{+}( \omega )$ is nonzero for $\omega > \omega_{\rm sp}$), 
but the following qualitative features are stable.  
(i) The mode $\omega_{+}( k )$ shows a gap between $0$ and 
$\omega_{+}( 0 )$, and the
difference $\delta\rho_{+}( \omega )$ is only due, 
for $\omega < \omega_{+}( 0 )$, to the subtracted 
isolated surface plasmon (dashed line).
Just at this
frequency, the mode density $\rho_+( \omega )$ jumps to a positive 
value. This behaviour is due to the quadratic shape of the lower 
band edge in $\omega_{+}( k )$. As $\omega \to \omega_{\rm sp}$, 
$\delta\rho_{+}( \omega ) > 0$ because $\omega_{+}( k )$ is shifted
upwards relative to $\omega_{0}( k )$ (the group velocity is smaller).
This mode is hence an `anti-binding one'~\cite{henkel:023808}.
(ii) The mode $\omega_{-}( k )$ has a 
linear dispersion for small $k$, and the difference in mode density
can be worked out as the positive quantity
$\delta\rho_{-}( \omega ) \propto \omega / \Omega_{\rm p} \propto
\omega L$ (dashed line).
This mode is hence anti-binding in this region as well.
(iii) Near the frequency $\omega_{\rm sp}$, the mode $\omega_{+}( k
)$ [$\omega_{-}( k )$] gives a repulsive [attractive] contribution to
the Casimir energy, respectively.  Summing over both modes yields a
repulsive or attractive result depending on $L$, because the relative
weight of the binding and anti-binding regions changes.
Coming back to 
frequency shifts, it is easy to see from the large $K$ 
expansion of Eqs.(\ref{eq:z-parametrization}) that the following inequality 
holds
\begin{equation}
\abs{\omega_-(\mathbf{k},L)-\omega_{0}(\mathbf{k})}>
\abs{\omega_+(\mathbf{k},L)-\omega_{0}(\mathbf{k})} \quad
(\mathbf{k}\gg\omega_{\rm p}/c),
\end{equation}
At short distance, the plasmonic Casimir energy (which is actually 
the total Casimir energy) is thus attractive, as is well known
(see also Eq.\ref{converg}).

Let us finally note that as one moves away from the large-$k$
regime, retardation becomes increasingly important. The change in 
sign of the plasmonic Casimir energy can thus be seen as well as
a consequence of the finite speed of light.

\subsection{Cutting the mode branch}
\label{a:modes}

Recently, there has been some discussion on the way to split the field
modes into photon-like and plasmon-like
parts~\cite{Bordag:2006,lenac:218901}.
We comment in this section on the numbers one can obtain
when the plasmonic mode $\omega_{+}( k )$
is segmented in a different way.
(The mode $\omega_{-}( k )$ is subject to no controversy.)
The main conclusion we draw 
from this discussion is that the large distance behavior is dominated 
by mode branches near the light line. In addition, the sign is 
sensitive to the chosen subtraction (renormalization), and it may
happen that under this procedure, a pure evanescent branch ends up 
being counted among photonic modes. We also suggest that the branch 
of the plasmonic mode $\omega_{+}( k )$ that enters the propagating 
sector is perhaps one of the best examples of Casimir repulsion due 
to a standing wave mode. Consider the corresponding pressure: it is
repulsive due to photons bouncing on the mirrors. The attractive
force for a perfect cavity arises, all things told, from the 
subtraction of a similarly repulsive pressure from a standing wave
mode continuum (reflected from the mirrors' backfaces).  Now, the
counterpart for the plasmonic mode is a single-interface evanescent
mode with zero pressure so that the repulsive force survives the
subtraction.

\begin{figure}[bth]
    \centerline{%
    \includegraphics*[width=70mm]{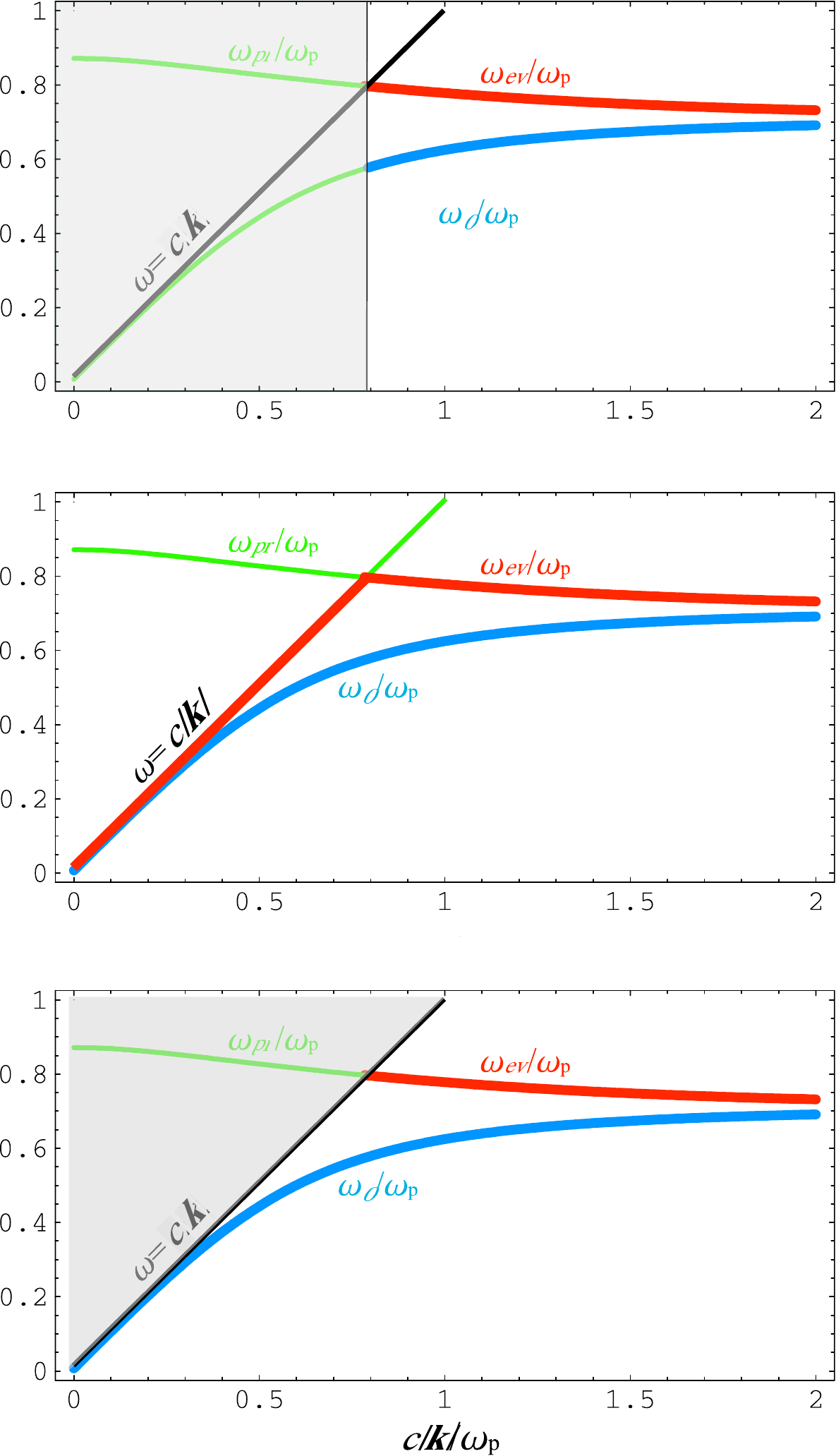}
    }
\caption[]{
Illustration of different segmentation of the plasmonic 
modes and the chosen renormalization. Thick
lines mark the segments that are taken into account in the different
approaches. We write
$\omega_{\rm pr}$ and $\omega_{\rm ev}$ for those parts of the
mode  $\omega_{+}( k )$ where the 
field between the mirrors is propagating or evanescent, respectively.
\\
Top: Bordag \cite{Bordag:2006}, the modes $\omega_{+,0}( 
k )$ (red, blue) start at the wavevector $k_{\rm c}$ 
where $\omega_{+}( k )$ reaches the light cone. 
Middle: one possibility suggested by the comment of Lenac
\cite{lenac:218901}. The mode
$\omega_{\rm ev}( k )$ is continued, for $0 \le k \le k_{\rm c}$, by the
light line $\omega = c k$ (red) and renormalized by the entire
branch of $\omega_{0}( k )$ (blue).
A particular splitting of the Lifshitz formula 
into propagating and evanescent modes turns out to yield the same result.
Bottom: another possibility compatible with Lenac's paper.
Only the evanescent branch $\omega_{\rm ev}( k )$ ($k \ge k_{\rm c}$, 
red) is taken into account and renormalized by 
$\omega_{0}( k )$ ($k \ge 0$, blue).}
\label{fig:mode-cuts}
\end{figure}

Bordag~\cite{Bordag:2006} is calling `plasmon mode' only the evanescent 
branch of $\omega_{+}( {k} )$ that exists for $k \equiv |{\bf k}| 
> k_{\rm c} \equiv
\omega_{\rm p} / (c \sqrt{ 1 + \Omega_{\rm p} / 2})$ 
(see Fig.\ref{fig:mode-cuts}, top). The segment within the light cone 
actually does not appear explicitly in Eq.(24) of 
Ref.\cite{Bordag:2006}, but is 
implicitly contained in the total Casimir energy (the
photonic contribution is computed by subtracting the plasmonic one). 
The evanescent segment
of $\omega_{+}( {k} )$ is renormalized by subtracting the isolated
surface plasmon, $\omega_{0} ( {k} )$, over the same range $k_{\rm c} < 
k < \infty$, as shown in Fig.\ref{fig:mode-cuts} (top). The 
range $0 < k < k_{\rm c}$ is left out (although it depends on $L$ via 
$k_{\rm c}$). This subtraction is sufficient to 
get a vanishing energy as $L \to \infty$ 
because $\omega_{+}( {k} ) \to \omega_{0}( {\bf k} )$
exponentially fast for $k > k_{\rm c}$. (In addition, $k_{\rm c} \to 0$.)
The integration over the branches chosen in Ref.\cite{Bordag:2006}
corresponds to the following correction factor to the Casimir energy:
\begin{eqnarray}
&& \eta_{\rm B} = \eta_{\rm L}
\nonumber\\
    && {}-\halfprefactor
    \Big[ 
    \int\limits_{0}^{z_{\rm c}}\!\sqrt{ g_{0}( z ) } dz
    + \frac23 \left( \Omega_{0}^{3}( K_{\rm c} ) 
    - K_{\rm c}^3 \right)
    \Big]
    \label{eq:Bordag-difference}
\end{eqnarray}
with
\begin{equation}
    \eta_{\rm L} = - \halfprefactor
    \int\limits_{0}^{\infty}\!
    \sum_{a = \pm,0} c_{a} \sqrt{ g_{a}( z ) } dz
    \label{eq:alternative-eta}
\end{equation}
Here, $K_{\rm c} = k_{\rm c}L$ and $z_{\rm c}$ solves the equation
$K_{\rm c}^2 = g_{0}( z_{\rm c} )$ (at this parameter value, the
dispersion relation $\omega_{0}( k )$ reaches $k = k_{\rm c}$).
We have checked that at short distance, this correction is negligible 
compared to the leading order $\eta_{\rm pl} \propto \omega_{\rm p} L$.
At large distance, however, the integrals in~(\ref{eq:Bordag-difference})
are both of order $\Omega_{\rm p}^{1/2}$ (see 
Section~\ref{s:large-distance}) and the difference 
$ \Omega_{0}^{3}( K_{\rm c} ) - K_{\rm c}^3$, too. Their contributions
come with different signs, leading in the end to a correction factor
that is attractive and scales like $\eta_{\rm B} \approx 
1.6240 \,\Omega_{\rm p}^{1/2}$ at large distance. 

A similar analysis can be done for the mode definition sketched in
Fig.\ref{fig:mode-cuts} (middle): the plasmonic mode is continued 
along the light line for $k < k_{\rm c}$ and renormalized by the 
\emph{entire} dispersion branch $\omega_{0}( k )$.   For $L \to
\infty$, as $k_{\rm c} \to 0$, the renormalized energy vanishes. 
The corresponding correction factor is given by $\eta_{\rm L}$ 
[Eq.(\ref{eq:alternative-eta})]
which does not contain any integrated term.  The short-distance
behaviour is the same as in the present paper, and at large distance,
we have $\eta_{\rm L}( L ) \approx - 1.6600 \Omega_{\rm p}^{1/2}$.
This corresponds to repulsion as with our convention, but with a
smaller numerical coefficient.  We argue below that this result can
also be obtained
by a splitting of the Lifshitz formula for the Casimir energy. Let us 
mention that if the segment $0 \le k \le k_{\rm c}$ of the light line 
is not taken into account (Fig.\ref{fig:mode-cuts}, bottom))
then the large distance behaviour 
shows an attractive term $\eta \propto \Omega_{\rm p}^{3/2}$. Both results
do not fit with the curves presented by Lenac~\cite{lenac:218901},
although our calculation tries to follow the spirit of his description.
It is not clear to us from his sparse description which renormalization 
scheme was used in the end.

The Lifshitz approach to the Casimir energy leads to the 
correction factor $\eta_{\rm L}$ as follows. We write the right-hand
side of Eq.(\ref{en}) in the equivalent form
\begin{equation}
    E_{\rm L} = - {\rm Im}\, \sum_{\mu,\mathbf{k}}\int\limits_{0}^{\infty }
    \frac{\mathrm{d}\omega }{2\pi}
    \hbar \omega 
    \frac{ d }{ d\omega }
    \log D_{\mu}[ \omega, {\bf k} ]
    \label{eq:Lifshitz-2}
\end{equation}
where the dispersion function
$D_{\mu}[ \omega, {\bf k} ]$ is defined in~(\ref{modes1}).
This expression has a structure very similar to the so-called 
``argument principle'' where the zeros (and poles) of the argument 
of the logarithm define the eigenfrequencies of the system (of the
reference system), respectively~\cite{most}, and each mode contributes
its zero point energy.  In other words, the imaginary part of the
logarithmic
derivative can be read as a density of modes (suitably renormalized). 
We isolate the contribution of evanescent modes by restricting the 
$\omega$-integration domain to $0 \le \omega \le c|{\bf k}|$ (so that
$\kappa = \sqrt{ |{\bf k}|^2 - \omega^2/ c^2 }$ is real as it should for 
evanescent waves). As discussed in Sec.\ref{s:real-cavity-modes},
zeros and poles of $D_{\mu}[ \omega, {\bf k} ]$
occur for the plasma model only for TM-polarized evanescent waves. 
A simple calculation leads to
\begin{eqnarray}
    && 1 - (r_{\mathbf{k}}^{TM})^{2} e^{- 2 \kappa L} =
    \left( 1 - (r_{\mathbf{k}}^{TE})^{2} e^{- 2 \kappa L} \right) 
    \times
    \label{eq:link-TM-dispersion-to-g-pm}
    \\
    &&
    \left( \frac{ g_{+}( \kappa^2 L^2 ) - (\omega L/c)^2
    }{ g_{0}( \kappa^2 L^2 ) - (\omega L / c)^2 }
    \right)
    \left( \frac{ g_{-}( \kappa^2 L^2 ) - (\omega L/c)^2
	}{ g_{0}( \kappa^2 L^2 ) - (\omega L/c)^2 }
	\right)
	\nonumber
\end{eqnarray}
where the functions $g_{\pm}( z )$ defined in~(\ref{eq:omega-plus},
\ref{eq:omega-minus}) appear.  The factor involving
$r_{\mathbf{k}}^{TE}$ shows no singularities for evanescent waves.
With the change of variable $k \mapsto z = (\kappa L)^2$, we see that
the two factors in the second line
of~(\ref{eq:link-TM-dispersion-to-g-pm}) have simple zeros (poles) at
the mode frequencies $\omega_{\pm}$ ($\omega_{0}$), respectively [see
Eq.(\ref{fpm})].  A calculation using the argument principle and the
symmetry property $\epsilon ( - \omega ) = \epsilon( \omega )$ for
lossless response functions, then leads straightforwardly to
Eq.(\ref{eq:alternative-eta}).

%


\section{Conclusion and Discussion}

In this paper we evaluate the contribution of plasmonic modes to
the Casimir force using the plasma model to describe the optical
response of the medium. Simple analytical expressions are found, in 
particular for the small and large distance asymptotics. We
introduced a correction factor $\eta_{\rm pl}( L )$ that gives the plasmonic
contribution to the Casimir energy, $E_{\rm pl}( L )$, in units of the 
Casimir energy $E_{\rm Cas}( L ) \propto -1/L^3$~[Eq.(\ref{ECas})].  It
turns out that $\eta_{\rm pl}( L )$ is small but positive at short
distance, correctly reproducing van Kampen's result~\cite{vankamp}.
Quite surprisingly, the plasmonic contribution changes sign at 
the fairly short
distance $L/\lambda_{\rm p}\sim 0.08$. For larger cavity lengths,
$\eta_{\rm pl}( L )$ becomes negative and leads to the unusual scaling 
$E_{\rm pl} \propto +L^{-5/2}$ as $L \to \infty$. This behaviour
clearly shows that the plasmonic modes
are much more important for the Casimir effect than usually
anticipated. They do not only dominate in the short distances limit,
but they also give a large repulsive contribution at large distances.

\begin{figure}[tbh]
 \center
   \includegraphics*[width=8cm]{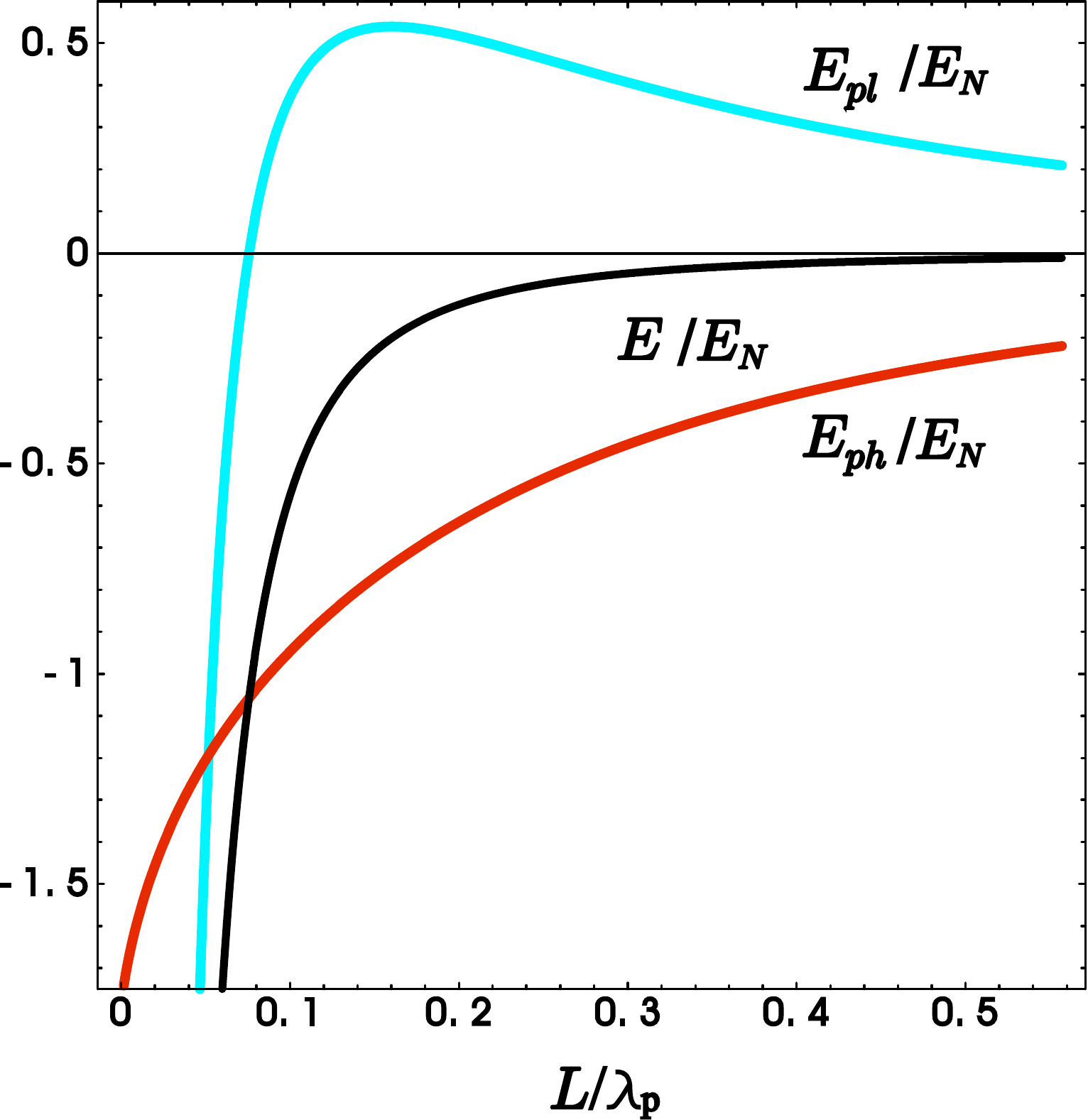}
   \caption{A plot of plasmonic $E_{\rm pl}$, photonic $E_{\rm ph}$ and 
total Casimir $E$ energy vs.\ distance $L$, normalized to the plasma
wavelength $\lambda_{\rm p}$. We normalize the energy to
  $(2\pi)^3\frac{\hbar c\pi^2A}{720 \lambda_{\rm p}^3}$.
  The plasmonic energy shows a maximum for 
$L \approx 0.16\,\lambda_{\rm p}$ (the corresponding force changes sign), 
while the photonic energy provides an attractive contribution at all
distances.}\label{nrgcont}
\end{figure}

We have calculated as well (see also~\cite{Bordag:2006}) 
the photonic mode contribution that turns out to be 
a monotonous function of the
distance $L$ (Fig.\ref{nrgcont}); it actually approaches a 
constant as $L \to 0$.
Its large distance behaviour contains as leading order a
negative $L^{-5/2}$ term that exactly cancels the plasmonic
contribution. The Casimir energy is thus
the balance of two contributions of 
equal magnitude which nearly cancel each other.


It would be interesting to investigate if a change in the
field-mirror coupling could somehow influence this detailed
balance and therefore the value or even the sign of the Casimir
force. This could be the case for
nanostructured surfaces, since the plasmonic modes are associated
with the electron charge density oscillations at the vacuum/metal
interface. This route has already been explored within a different
context, that of metamaterials in the visible frequency range.
It has been shown that arrays of metallic dots or rods 
\cite{Panina:2002,Podolskiy:2002,Grigorenko:2005} exhibit a strong magnetic
response in the visible, including a band with
negative magnetic permeability. This behaviour arises again from 
plasmon modes: they are here concentrated on the metallic particles
and their characteristics can be 
tuned with the particle shape. In the array, the plasmons delocalize 
and lead to a resonant electric and magnetic response.
A significant modification of the Casimir force, even a change in sign, 
could be realistic with these materials~\cite{Henkel:2005}.

The limits of validity of our results are imposed by the
applicability of the plasma model. The approach is not intended to make
quantitative predictions because the response of intraband transitions 
in real metals would require a more complicated dielectric function.
We are also restricted to fairly 
short distances where the relevant mode frequencies are sufficiently 
large compared to the dissipation rate. 
From Fig.\ref{nrgcont}, 
one can see, however, that the most striking 
effects due to plasmonic modes (change in sign 
and near cancellation of plasmonic and photonic contributions)
indeed appear at short distances,  $L \le \lambda_{\rm p}$, where the 
lossless plasma response is a suitable approximation.
We therefore believe that at least for this
range of distances our results are quite generally valid.

In our description, the main responsible for a repulsive Casimir 
interaction is the
plasmonic mode $\omega_{+}$. This mode
crosses the border between the evanescent sector
and the propagative sector, and we have it considered as being
completely part of the `plasmonic' set of
modes~\cite{intravaia:110404}.  This is an `adiabatic' definition that
is strongly suggested by the continuous change in the mode function
plotted in Fig.\ref{opm}. Other splittings into evanescent and 
photonic modes have been applied in the
literature~\cite{Bordag:2006,lenac:218901}, and we have given a
brief review in Section~\ref{a:modes}.
The total Casimir energy is of course immune to these wordings.
However, if one considers a structured surface,  
the mode branch $\omega_{+}$ will change \textit{as a whole}, and
by analyzing this change, one could easily predict the corresponding
modification of the Casimir energy.


\acknowledgments

F. I. is grateful to M. Sokolsky for providing a critical reading of
the paper. F.I. enjoyed financial support from Fondazione Angelo 
della Riccia, from QUDEDIS, a scientific program of
the European Science Foundation (ESF), and from FASTNet, a Research
Training Network within the 5th Programme Framework of the European
Union. F.I. and A.L. thank S. Reynaud for useful discussions.

\appendix

\section{Plasmonic dispersion relations}
\label{s:dispersion-relations}

We collect in this appendix some properties of the solutions
$\omega_{\pm}( {\bf k} )$ of the dispersion relation 
Eq.(\ref{modes1}) in the evanescent sector.  Similar discussions can 
be found in Refs.\cite{Economou69,Chang73}.

To simplify the notation, we use the dimensionless variables
\begin{equation}
    K = \abs{\mathbf{k}}L, \quad \Omega = \omega L / c,
\label{eq:scaled-variables}
\end{equation}
and write the ratio $L / \lambda_{\rm p} = \Omega_{\rm p} / 2\pi$. 
By symmetry, the mode frequencies do not depend on the angle of 
$\mathbf{k}$ within the mirror plane.
The zeros of Eq.(\ref{modes1}) and can be split into  `odd' and `even' 
cases 
\begin{equation}
\begin{cases}
1 - r^\mu_{K}[ \Omega ] e^{- \kappa L} = 0\\
1 + r^\mu_{K}[ \Omega ] e^{- \kappa L} = 0
\end{cases}
\end{equation}
For the TM-polarization, the 
solutions to these equations
can be written in the parametrized form
\begin{subequations}
\label{eq:z-parametrization}
    \begin{equation}
\label{fpm}
 \left\{
 \begin{array}{rcl}
 K_{\pm} &=& \sqrt{ z + g_{\pm}(z) }
 \\
     \Omega_{\pm} &=& \sqrt{ g_{\pm}(z) }
 \end{array}
 \right.
\end{equation}
 where
\begin{gather}
 g_+( z ) = \frac{ \Omega^2_p\  \sqrt{z} }
{  \sqrt{z}  + \sqrt{ z + \Omega_{\rm p}^2 } 
\tanh(\textstyle\frac12\sqrt{ z })}
\label{eq:omega-plus}
\\
g_-(z) = \frac{ \Omega^2_p\   \sqrt{z}  }
{ \sqrt{z} + \sqrt{  z  + \Omega_{\rm p}^2} 
\coth(\textstyle\frac12\sqrt{ z })}
\label{eq:omega-minus}
\end{gather}
The parameter $z = (\kappa L)^2$ runs over a part of the real axis
such that $z + g_{\pm}(z) \ge 0$ in Eq.(\ref{fpm}): it is positive in 
the evanescent sector and negative in the propagating one. The square
root is chosen with ${\rm Im}\,\sqrt{z} < 0$.
For completeness, we give here the form of
$g_{+}( z )$ with negative argument:
\begin{equation}
    g_{+}(  -z' ) = \frac{ \Omega^2_p\  \sqrt{ z'} }
{  \sqrt{z'}  + \sqrt{ \Omega_{\rm p}^2 - z'} 
\tan(\textstyle\frac12\sqrt{ z' })} ;
    \label{eq:g-plus-neg-argument}
\end{equation}
it takes positive values as long as $z'$ is not too large (see
Eqs.(\ref{eqy},\ref{eq:zplus-limits}) below).

%
%
%
%
%

We shall also need the surface plasmon dispersion for a single 
metal-vacuum interface, $\Omega_{0}( K )$.
It can formally be found by taking the limit
$L \to \infty$ (while $\kappa > 0$) in 
Eqs.(\ref{eq:omega-plus},\ref{eq:omega-minus}), leading to 
the function
\begin{equation}
    g_{0}( z ) = \frac{ \Omega^2_p \sqrt{z} }{
    \sqrt{z} + \sqrt{z + \Omega_{\rm p}^2} }
    \label{eq:g0-single-interface}
\end{equation}
\end{subequations}
that parametrizes according to the scheme of Eqs.(\ref{fpm}) the 
poles of $r^{TM}_{K}[ \Omega ]$. We note for completeness the
explicit expression for the single interface
\begin{equation}
\Omega_{0}( K ) =
\sqrt{ \frac{\Omega^2_p + 2 K^{2} 
- \sqrt{ \Omega^4_p + 4 K^4 }}{2}}
\end{equation}
(This is of course independent of $L$ since the scale factor $1/L$ 
can be removed throughout.)

In the evanescent sector, both functions
$z + g_{\pm}( z )$ are monotonous
so that exactly one evanescent solution is found for a given $K$. 
The limit $z \to \infty$ corresponds to large wave vectors $K$ where
we have 
\begin{equation}
\Omega_{a}( K ) \xrightarrow{K \to \infty}
\Omega_{\rm p} / \sqrt{ 2 },
\quad a = \pm, 0
.
\label{eq:large-K-expansion}
\end{equation}
The two surface plasmons decouple in this limit. A more detailed 
analysis of the asymptotic behaviour is given in Sec.\ref{s:plasmonic-Casimir}.

The modes $\Omega_{-}( K )$ and $\Omega_{0}( K )$ remain in the 
evanescent sector for $K \to 0$: analyzing the limit $z \to 0$, one 
gets the linear dispersions
\begin{eqnarray}
    \Omega_{-}( K ) &\approx& \frac{ K }{ \sqrt{ 1 + 2 / 
    \Omega_{\rm p} } }
\\
    \Omega_{0}( K ) &\approx& K
\end{eqnarray}

The mode $\Omega_{+}( K )$ crosses the light line when $z = 0$,
corresponding to the frequency $\Omega_{+0} = [g_{+}(0)]^{1/2} = 
\Omega_{\rm p} / \sqrt{ 1 + \Omega_{\rm p} / 2 }$. The 
dispersion relation enters the cavity sector in a
continuous and differentiable way. The same is true for the associated
mode function, as illustrated in Fig.\ref{opm} (inset).
It reaches $K = 0$ at a parameter
value $z = - z_{+}$ that is the solution of $z + g_{+}( z ) = 0$.
This can be simplified to the implicit equation
\begin{equation}
\label{eqy} 
\sqrt{ z_{+} } = \Omega_{\rm p} \cos(\textstyle\frac12\sqrt{ z_{+} })
\end{equation}
from which one gets the exact inequalities $0 \le z_{+} \le 
\min( \Omega_{\rm p}^2, \pi^2 )$
and the asymptotics
\begin{equation}
    z_{+} \rightarrow
\begin{cases}
\Omega_{\rm p}^2 & \text{for } \Omega_{\rm p} \ll \pi
\\
\pi^2 & \text{for } \Omega_{\rm p} \gg \pi
\end{cases}
     \label{eq:zplus-limits}
\end{equation}
In the first case, the mode ends at the bulk mode continuum 
(Fig.\ref{opm} is close to this situation).  The second case implies
that at large distance ($L \gg \lambda_{\rm p}$), the mode
stays close to the light line, as in Fig.\ref{disptm}.  One can also
calculate that the generic behaviour of  the dispersion relation for 
small $K$ is quadratic: 
\begin{equation}
    \Omega_{+}( K ) \approx 
    \Omega_{+}( 0 ) + \frac{ K^2 }{ 2 \Omega_{+}( 0 ) }
    \frac{ \sqrt{ \Omega_{\rm p}^2 - z_{+} } - 2 
	}{ \sqrt{ \Omega_{\rm p}^2 - z_{+} } + 2 } 
    .
    \label{eq:quadratic-dispersion}
\end{equation}



%
%



\end{document}